\PassOptionsToPackage{bookmarks=false}{hyperref}

\documentclass[sigconf]{acmart}
\usepackage[bookmarks=false]{hyperref}
\usepackage{amsmath,amssymb,amsfonts}
\usepackage{algorithmic}
\usepackage{graphicx}
\usepackage{textcomp}

\begin{document}

\title{Modeling Techniques for Logic Locking}

\author{Joseph Sweeney}
\email{joesweeney@cmu.edu}
\affiliation{%
  \institution{Department of Electrical and Computer Engineering,\\
  Carnegie Mellon University}
  \city{Pittsburgh}
  \state{PA}
  \postcode{15232}
}
\author{Marijn J. H. Heule}
\email{marijn@cmu.edu}
\affiliation{%
  \institution{Department of Computer Science,\\
  Carnegie Mellon University}
  \city{Pittsburgh}
  \state{PA}
  \postcode{15232}
}
\author{Lawrence Pileggi}
\email{pileggi@cmu.edu}
\affiliation{%
  \institution{Department of Electrical and Computer Engineering,\\
  Carnegie Mellon University}
  \city{Pittsburgh}
  \state{PA}
  \postcode{15232}
}
\settopmatter{printacmref=false}

\begin{abstract}
Logic locking is a method to prevent intellectual property (IP) piracy. 
However, under a reasonable attack model, SAT-based methods have proven to be  powerful in obtaining the secret key. 
In response, many locking techniques have been developed to specifically resist this form of attack.
In this paper, we demonstrate two SAT modeling techniques that can provide many orders of magnitude speed up in discovering the correct key.
Specifically, we consider relaxed encodings and symmetry breaking.
To demonstrate their impact, we model and attack a state-of-the-art logic locking technique, Full-Lock. 
We show that circuits previously unbreakable within 15 days of run time can be solved in seconds. Consequently, in assessing the strength of any given locking, it is imperative that these modeling techniques be considered. 
To remedy this vulnerability in the considered locking technique, we demonstrate an extended version, logic-enhanced Banyan locking, that is resistant to our proposed modeling techniques. 

\end{abstract}



\renewcommand\footnotetextcopyrightpermission[1]{} 

\acmConference[]{}{}{}

\maketitle

\section{Introduction}
Due to prohibitively high research and development costs, only a few foundries are manufacturing integrated circuits (ICs) in advanced technology nodes. Consequently, many IC companies tend to operate fabless, relying on untrusted foundries to manufacture their designs. Once a circuit is sent for fabrication, the foundry gains full visibility of the design in netlist form with minimal effort, allowing IP theft. This threat undermines the significant cost associated with developing digital circuits and is a growing concern in the IC industry \cite{Guin2017AOverproduction}.

To combat IP theft, a variety of logic locking techniques have been developed. These techniques add programmable elements to the logic of a digital IC. When programmed incorrectly, the elements disrupt the circuit, obfuscating the true functionality. The key, which correctly programs the elements, is stored in an on-chip, tamper-proof memory. This key is set post-manufacture, so the correct functionality is never revealed to the untrusted foundry.

Early examples of logic locking techniques inserted keyed exclusive-or (XOR) and multiplexer (MUX) gates to corrupt the next-state logic \cite{J.A.Roy2008EPIC:Circuits,Rajendran2015FaultEncryption}. Unfortunately, these methods have been largely broken using a variety of attacks, the most successful of which are miter-based SAT attacks \cite{Subramanyan2015EvaluatingAlgorithms}. Researchers have attempted to increase the difficulty of the miter-based attack by inserting resistant logic blocks into the locked circuit. 

These resistant locking techniques generally fall into two categories based on how they resist the miter-based SAT attack. The first group \cite{Xie2019Anti-SAT:Locking,Yasin2016SARLock:Locking,YasinWhatLocking,YasinProvably-SecurePractice} focuses on reducing the number of keys ruled out per attack iteration, significantly increasing the expected \textit{number of iterations}. In practice however, these techniques are susceptible to removal attacks since the circuitry is typically traceable through properties such as signal probability or Boolean sensitivity \cite{Yasin2017SecurityAnti-sat,sensitivity}. 

The second group \cite{Kamali2019Full-Lock,Shamsi2018Cross-Lock:Architectures} tries to extend the \textit{time per iteration}. This is done by adding SAT-hard instances into the circuit. These instances typically have many interdependent keys; a prototypical example is the lookup table (LUT) combined with configurable routing. The resulting locks resemble field-programmable gate arrays (FPGAs) embedded into the circuit and have been shown to be highly resistant to the current miter-based SAT attacks. 

In this paper, we explore the use of two modeling techniques targeting locked circuits from this second group.
These techniques are shown to be powerful tools in revealing the key, dramatically reducing attack run time. 
Specifically, the contributions of this work are the following:
\begin{itemize}
\item Proposal of relaxed encoding and symmetry breaking as modeling techniques for attacking locked circuits 
\item Demonstration of impact of the modeling techniques in attacking a state-of-the-art scheme, Full-Lock
\item Logic-enhanced Banyan locking, an improved version of Full-Lock, not susceptible to these new attack techniques
\end{itemize}

\section{Background}
\subsection{Attack Model}
In the characterization of the security of a locking technique, an attack model is used to specify assumptions regarding the adversary's ability. In this paper, as in all the aforementioned locking techniques, it is assumed that the adversary has access to two artifacts: the locked circuit's netlist and an unlocked version of the circuit. The unlocked circuit has the correct key set in its tamper-proof memory, affording the attacker black-box access, commonly referred to as an oracle. These artifacts correspond to the access a foundry is likely to have when manufacturing a commercial design. The netlist can be easily reversed engineered from the design data and the unlocked circuit can be obtained on the open market. It is also assumed that the adversary has access to the unlocked design's scan chains. While additional side-channel techniques may augment an attacker, they are considered outside the scope of the paper. 

In general, the problem being solved by the attacker is as follows. The attacker has access to an unlocked circuit containing a set of Boolean functions. Each function, $f:\{0,1\}^{n+k}\rightarrow \{0,1\}$, has $n$ normal inputs and $k$ unknown, fixed key inputs. The attacker also has knowledge of the structure of $f$. Using the unlocked circuit and knowledge of the structure, the goal of the attacker is to obtain a functionally equivalent version of the circuit.

\subsection{Propositional Satisfiability}
A common approach to deal with hard combinatorial problems, such as finding the key of locked circuits, is to encode them
into propositional logic and to solve the resulting propositional formulas with a satisfiability (SAT) solver. The performance of SAT solvers improved significantly in the last two decades and they are used for many applications in hardware and software verification~\cite{Biere:1999,software}. In recent years, SAT solvers have also been successfully applied to various attacks, such as hash collisions~\cite{SHA1-SAT} and mathematical challenges~\cite{EDP-SAT}.

The typical encoding of the SAT problem is in the conjunctive normal form (CNF). This form consists of a set of clauses that must all be satisfied. Each clause is a disjunction of literals. 
A circuit can be encoded into propositional logic via the Tseitin transformation \cite{tseitin1983complexity}. This transformation can take a circuit netlist and produce a set of clauses which, when collectively satisfied, will correspond to the original circuit's behavior. 

The most successful class of SAT solvers are based on the conflict-driven clause learning (CDCL) algorithm \cite{bayardo1997using,marques1999grasp}. Briefly, CDCL solvers work by repeatedly selecting a variable through heuristics and assigning a value. 
Implications from the assignments are determined using a highly optimized process called unit propagation. If a conflict is found, a clause is added to the formula that rules out assignments causing the conflict. Then the solver non-chronologically backtracks based on the conflict and continues, repeating this process until a solution is found or the problem is found to be unsatisfiable. 

\subsection{Miter-Based SAT Attack}
\begin{figure}[t]
  \centering
  \includegraphics[width=\columnwidth]{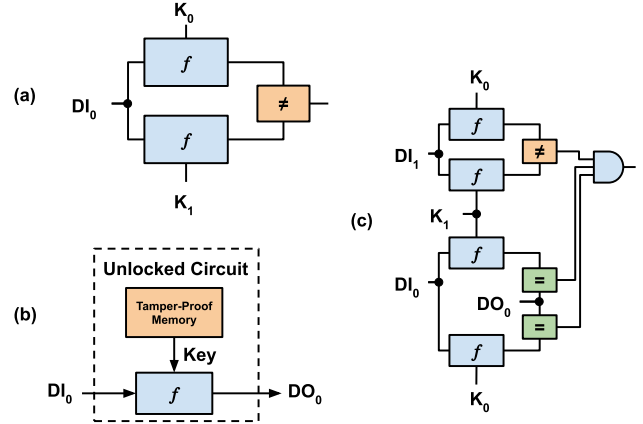}
  \caption{Miter-based SAT attack steps: (a) Miter circuit construction, (b) Unlocked (oracle) circuit produces correct IO functionality (c) Addition of learned IO constraint to miter circuit}
  \label{fig:sat}
\end{figure}

The above attack model enables the mounting of a more targeted, miter-based SAT attack. This attack uses the netlist and unlocked circuit to iteratively produce input-output (IO) relationships \cite{Subramanyan2015EvaluatingAlgorithms}. These relationships are used to rule out all keys that do not produce the same behavior, narrowing the space of possible circuit functionalities. The IO relationships are efficiently learned through a three-step procedure: \textbf{I.} First, a miter circuit~\cite{miter} is used to determine an input that is guaranteed to rule out at least a single key. A miter circuit consists of two copies of the original circuit with the inputs tied together, the key inputs kept separate, and the outputs connected to comparators. A diagram of the connections is shown in Fig. \ref{fig:sat}a. Additional key constraints, such as timing and loop breaking, can be conjuncted with the miter output. A SAT solver is used to find a setting of the shared input (I) and key inputs ($\mathrm{K_0},\mathrm{K_1}$) such that the output of the miter circuit is logic 1. By construction, the solution to the SAT problem will have two different keys that, at that input value, disagree on the output value. The shared input value found by the solver is termed a differentiating input (DI). \textbf{II.} Next, as depicted in Fig. \ref{fig:sat}b, the learned DI is applied to the oracle circuit to determine the differentiating output (DO), forming an input-output (IO) pair which the correct key must respect; any key that does not conform to this IO pair is incorrect. \textbf{III.} Finally, as shown in Fig. \ref{fig:sat}c, the IO  pair is added as a constraint to the miter circuit for the next iteration. Now, any keys that satisfy the miter circuit will also satisfy the learned IO relationship. While each relationship is guaranteed to rule out at least one key, in practice, a larger portion of the key space is ruled out due to overlapping key functionalities at a given input. These steps repeat, adding more constraints until the miter circuit is unsatisfiable. At this point, any key that respects all learned IO relationships will be a functionally correct key. 

\subsection{Full-Lock}

Full-Lock is a logic locking technique specifically developed to be resistant to the miter-based SAT attack \cite{Kamali2019Full-Lock} via increasing the execution time of each iteration. This is done by integrating SAT-hard logic into the circuit using a combination of routing obfuscation and look-up tables (LUT). The added logic is highly symmetric with many keys mapping to the same functionality. Symmetry is known to be difficult for SAT solvers, trapping the algorithm by spending time exploring solutions that are isomorphic \cite{crawford1996symmetry}. Furthermore, unit propagation of the circuit is hindered as each configuration depends on many keys: in order to determine the output of the Full-Lock circuitry, most keys must be assigned. Finally, the obfuscation is parameterized such that locking scheme's clauses to variables ratio is close to 4.26, the phase-transition density for uniform random 3-SAT (SAT instances with exactly 3 variables per clause)~\cite{ding2015proof}. 
Intuitively, instances with a higher ratio are over-constrained making contradictions easier to find and those with a lower ratio are under-constrained with potentially many satisfying solutions. While the instances produced by Full-Lock are not uniform random 3-SAT, and therefore likely have a different optimal ratio, the locking still produces hard SAT instances.

\begin{figure}[t]
  \centering
  \includegraphics[width=\columnwidth]{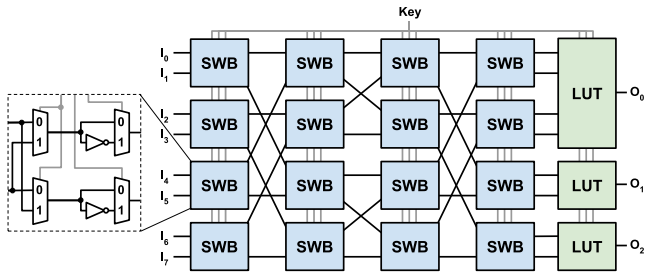}
  \caption{Full-Lock diagram. Each LUT replaces a gate from the original circuit; the switch boxes permute and invert their input signals.}
  \label{fig_fulllock_diag}
\end{figure}

Full-Lock utilizes configurable routing and LUTs to obfuscate a set of gates and their corresponding input connections. The configurable routing is implemented with Banyan networks, a class of logarithmic networks, that permutes connections based on a key~\cite{Kamali2019Full-Lock}. The network is made up of a series of 2-input switch boxes which connect the inputs to the outputs, either directly passing through or switched. Additionally, Full-Lock adds the ability to invert the polarity of the signals in each switch box. 
Diagrams of the switch boxes and overall network are shown in Fig. \ref{fig_fulllock_diag}. 
The specific Banyan network configuration used has $2*log_2(N)-2$ stages where $N$ is the network's input width (equal to the number of permuted lines). 
The Banyan network is almost non-blocking, meaning that almost all input to output connection permutations are possible. 

The locking procedure is as follows. A set of gates with the desired number of total inputs is randomly selected from the circuit. A Banyan network is inserted into the circuit.
The nets fanning into the selected gates are randomly inverted and connected to the network's inputs. The selected gates are replaced with LUTs of appropriate input size. 
The outputs of the network are connected the LUT inputs such that under the correct key, each LUT will receive the original inputs with proper polarity. The key to the circuit is thus the concatenation of the LUT and network configuration bits. The random selection of gates opens the possibility for combinational loops to be formed in the circuit. This has no impact on the circuit when the correct key is applied as all feedback paths will be broken. However, if not ruled out, these loops will corrupt the miter-based SAT attack. Several methods of building loop-breaking key constraints have been developed to re-enable the attack \cite{Zhou2017CycSAT:Encryptions, shamsi2019icysat}. 
 
As is, this locking method appears resistant to the miter-based SAT attack. 
The authors of the original work ran the attack for 15 days without termination on instances with 32 circuit lines permuted. Additionally, the authors considered a removal attack. The added circuitry is easily identifiable, even after synthesis due to the key lines and regular structure. Despite this, Full Lock is also resistant to a removal attack as the selected gates have been replaced with LUTs and the correct interconnections and polarities of their inputs are unknown. 

\section{Modeling Techniques}
Critical to the performance of SAT solvers is the encoding of the problem. Many problems become hard for SAT due to a poor encoding.
Often the best encoding is found after trying several different strategies \cite{biere2009handbook}.
Thus, when assessing the security of a locking technique, the encoding used can drastically influence the results. In this section, we describe two modeling techniques that are widely applicable to logic locking. We demonstrate the application of each technique to the example locking method, Full-Lock. 

\subsection{Relaxed Models}
\label{section_relaxed_models}
Each iteration of the miter-based SAT attack satisfies the miter circuit while respecting the system model.
The system model captures the potential behaviors of the locked circuit under different keys and is encoded into propositional logic allowing the SAT solver to generate meaningful inputs. 
However, the exact system model can be difficult to specify or too complex for SAT solvers to efficiently handle. Often, a close analog to the original behavior can be captured with a much simpler encoding. Substituting the system model can allow significant decreases in attack time, sacrificing precision for reduced complexity. 

Several factors must be considered when building a relaxed model for a locked circuit. 
\textit{First}, the model's variables do not all need to directly map to system's logic. In fact, the only requirement on the variable mapping is that the inputs and outputs remain directly mapped between the encoding and original system model so that the produced DIs can be run on the oracle and the resulting DI-DO pair can be added to the miter. \textit{Next}, the relaxed model must be able to produce a super-set of the input-output relationships under all key values. Perhaps counter intuitively, specifying a super-set of behaviors can be easier than the exact set. \textit{Finally}, while the key variables do not need to be directly encoded, there must be a mapping from the relaxed model back to a valid key configuration of the original system. 

An example of relaxed modeling is seen in TimingSAT \cite{Li2019TimingSAT:Obfuscation}, an attack methodology for TimingCamouflage \cite{8341985}. TimingCamouflage substitutes flip-flops with combinational logic delays. This disrupts a naive attack strategy because a reverse engineered netlist will be missing flip-flops that correspond to the correct functionality. It is assumed that to obtain the system functionality, an attacker must meticulously time the circuit and check all possible paths for potential combination logic delays replacing a flip-flop. However, TimingSAT simply substitutes a relaxed model, overestimating the possible locations where a combinational delay may be used as a flip-flop. In each potential flip-flop location, a MUX is inserted selecting between a flip-flop or wire. The functionality is then determined using the standard miter-based SAT attack, solving for the proper MUX settings. 

A relaxed encoding can also be used to remove key interdependence. Often the functionality of a locked circuit will depend on a large portion of the keys. To determine the output for a given input, the SAT solver must branch on many of the key variables. However, in some cases the functionality can be separated from the key variables. This allows the functionality to be selected without assigning all keys. An analogous example is encoding integers. The typical circuit for handling integers is representing them with binary numbers, however, to select an integer value all variables representing the binary number's bits must be assigned. For SAT solvers, an often more efficient strategy is one-hot encoding. Here, a value can be directly assigned by setting a single variable true (and unit propagating the others to false). In a similar sense a circuit functionality can be decoupled from the key bits, directly selecting the functionality rather than assigning all key bits. 

\begin{figure}[t]
  \centering
  \includegraphics[width=.5\columnwidth]{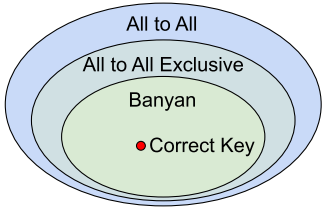}
  \caption{Relaxed models for Banyan network}
  \label{fig_mod_diag}
\end{figure}
\begin{figure}[t]
  \centering
  \includegraphics[width=.8\columnwidth]{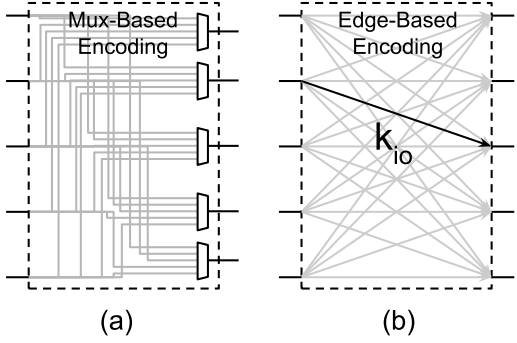}
  \caption{Edge-based and MUX-based encoding schemes for the all-to-all model}
  \label{fig_enc_diag}
\end{figure}
Using this relaxed encoding strategy, we consider our example technique, Full-Lock. 
As previously established, the Banyan network is a SAT-hard circuit due to its large amounts of symmetry, key interdependence, and poor unit propagation behavior. Despite its complexity, the functionality is very simple: the outputs of the network are a permutation of the inputs. Due to the structure of the network, some permutations are prohibited, and others can be selected by multiple key settings. If we relax the encoding of the network, allowing the prohibited permutations in our model, we can significantly reduce the complexity. 

We consider two relaxed models in place of the Banyan network: \textit{all-to-all}, wherein every input can be routed to every output, and \textit{all-to-all exclusive}, which additionally restricts an input to be routed to only a single output. 
A diagram of these functionalities is shown in Fig. \ref{fig_mod_diag}. The correct key is in the set of functionalities that the Banyan network allows, which is a subset of the all-to-all exclusive model functionalities, and in turn, the all-to-all model functionalities.

From a circuit designer’s perspective, the natural way to encode all-to-all functionality uses an N-to-1 MUX for each output, similar to the structure depicted in Fig. \ref{fig_enc_diag}a. This can be easily specified in a high-level language such as verilog, then synthesized to a gate-level representation. The Banyan network in Full-Lock can then be substituted for these gates.
Just as in the typical miter-based SAT attack, the circuit can then be encoded into SAT via the Tseitin transformation. The all-to-all exclusive encoding can be formed in the same fashion, adding circuitry to ensure that the select bits of each MUX are different.

We also consider an edge-based strategy in which a key variable, $k_{io}$, is created for each possible input to output connection. A diagram of this encoding is depicted in \ref{fig_enc_diag}b. The CNF of the encoding is shown below where $x_j$ is a variable representing a net $j$, $I$ is the set of nets fanning into the Banyan network, and $O$ is the set of nets in its fanout.  
\begin{equation}
\bigwedge_{i\in I, o \in O}k_{io} \to (x_i\leftrightarrow x_o)
\label{eq_a2a}
\end{equation}
To ensure proper functional behavior we must also enforce that each network output is only connected to one input. This can be done using a cardinality encoding over the same variables as below:
\begin{equation}
\bigwedge_{o\in O} ExactlyOne(\{k_{io}:i\in I\})
\label{eq_a2ao}
\end{equation}
The edge-based all-to-all exclusive encoding is created with the additional clauses:
\begin{equation}
\bigwedge_{i\in I} ExactlyOne(\{k_{io}:o\in O\})
\label{eq_a2ae}
\end{equation}

Running the miter-based SAT attack on these encodings will produce the correct mapping from the network inputs to outputs. Obtaining the corresponding key for the original system model can be done by finding a key that propagates the same paths in the Banyan network. 
Our models allow a greater function space, but with an encoding much more amenable to SAT solvers as we will see in section \ref{sec_rel_enc}. 

\subsection{Symmetry Breaking}
\label{sec_sym}
Another modeling technique that is not entirely exclusive from relaxed encodings, but can be applied on its own, is symmetry breaking. 
In the context of SAT, a symmetry is defined as a permutation of variable assignments which maps one solution onto another~\cite{shatter}. In the miter-based SAT attack, symmetry results from classes of keys producing the same circuit functionality. All equivalent keys will be equisatisifiable with respect to the miter circuit inputs. If symmetry exists in the locked circuit, the attack will waste time exploring isomorphic parts of the search space. 

Symmetry breaking in the miter-based SAT attack context entails ruling out all but one key from each equivalence class. Ideally, this is done with minimal additional clauses being added to the problem, otherwise the additional problem complexity may outweigh any benefit.
While not specifically labeled as symmetry breaking, this strategy has been utilized in the key-sensitization attack on Strong Logic Locking \cite{7362173}, wherein back-to-back key XOR gates are converted into a single XOR. 

Several examples of symmetry are seen in Full-Lock. In the Banyan network, multiple keys produce the same permutations of the inputs on the outputs. Additionally, the keys which optionally invert the switch box outputs are highly symmetric: all configurations of these keys can be reduced to a single bit for each output specifying whether it is inverted. 
Our relaxed models of the network already remove these two symmetries.
However, there remains a significant amount of symmetry in Full-Lock's LUTs.

By themselves, LUTs have no symmetric assignments, but when coupled with external circuitry, they can become highly symmetric. This property is good for Field Programmable Gate Arrays (FPGAs) wherein flexible configurations can help meet timing, power, and area constraints, however, this flexibility hinders the miter-based SAT attack as the same logical functionality can be specified many ways. 

Full-Lock allows the LUT-inputs to be permuted which creates \textit{LUT configuration, input permutation} pairs that are symmetric.
In Table \ref{tab_sym}, we show all the symmetric configurations of a 2-Input LUT with the inputs permuted. Each group with more than one equivalence is highlighted in a different shade of blue. Within the highlighted groups, permuting the inputs allows a single LUT configuration to function equivalently to the others. Thus, only one LUT configuration per group is needed. 
In the 2-input LUT case, only 4 out of 16 configurations are eliminated, however, as the input width increases the number of symmetric configurations grows significantly. For a 4-input LUT, there is over an order of magnitude reduction in the number of remaining configurations. 
\begingroup
\setlength{\tabcolsep}{5pt} 
\begin{table}[t]
\caption{2-Input LUT Symmetries Under Permuted Inputs}
\definecolor{g0}{HTML}{2e59a8}
\definecolor{g1}{HTML}{5d8abd}
\definecolor{g2}{HTML}{8abccf}
\definecolor{g3}{HTML}{c5eddf}

\newcolumntype{a}{>{\columncolor{g0}}c}
\newcolumntype{b}{>{\columncolor{g1}}c}
\newcolumntype{d}{>{\columncolor{g3}}c}
\newcolumntype{e}{>{\columncolor{g2}}c}
\begin{center}
 \resizebox{\columnwidth}{!}{%
\begin{tabular}{ c|ccab abcc cced edcc}
    $K_0$ ($I_1,I_0=0,0$) &0 & 1 & 0 & 1 &   0 & 1 & 0 & 1 &  0 & 1 & 0 & 1 &  0 & 1 & 0 & 1\\
    $K_1$ ($I_1,I_0=0,1$) &0 & 0 & 1 & 1 &   0 & 0 & 1 & 1 &  0 & 0 & 1 & 1 &  0 & 0 & 1 & 1\\
    $K_2$ ($I_1,I_0=1,0$) &0 & 0 & 0 & 0 &   1 & 1 & 1 & 1 &  0 & 0 & 0 & 0 &  1 & 1 & 1 & 1\\
    $K_3$ ($I_1,I_0=1,1$) &0 & 0 & 0 & 0 &   0 & 0 & 0 & 0 &  1 & 1 & 1 & 1 &  1 & 1 & 1 & 1\\
\end{tabular}}
\label{tab_sym}
\end{center}
\end{table}
\endgroup

Full-Lock's input permutation symmetry can be broken by enforcing an ordering on the inputs connected to each LUT. This ensures that for every combination of inputs routed to a LUT, only a single permutation is allowed, ruling out all unnecessary configurations. To create this ordering, we add a unary mapping of the key variables of our edge-based encoding. For each LUT, where $s_{io}$ is an auxiliary variable representing the unary mapping for a key $k_{io}$ and $O_L$ is the ordered set of network outputs that connect the LUT, we add the clauses in Eq. \ref{eq_order} to our solver. 

\begin{equation}
\begin{aligned}
\bigwedge_{i\in I,o\in O_L} & (k_{io}\to s_{io})\wedge (k_{io}\to \neg s_{io+1})\\
& \wedge (k_{io}\to s_{i+1o}) \wedge (s_{io}\to \neg s_{i+1o})
\label{eq_order}
\end{aligned}
\end{equation}

\section{Logic-Enhanced Banyan Locking}
\subsection{Overview}
Based on our attack data in Section \ref{sec_rel_enc} and the results from the original Full-Lock work, it is clear the Banyan network structure creates an instance that is difficult for the miter-based SAT attack. The strengths of the network are the large number of cycles it can potentially create, the interdependence of keys, and the lack of intermediate outputs. 
However, with the proposed modeling techniques, we have exposed holes in the original formulation. Here, we describe a remedy based on breaking the assumptions of the modeling techniques through the addition of logic internal to the network. 

Our locking technique, logic-enhanced Banyan locking, uses the same Banyan structure as Full-Lock, however, the functionality is extended beyond the simple invert and permute. This is achieved by moving logic from the locked circuit into the switch boxes of the Banyan network. 
In the original Full-Lock switch box, two key bits are used to optionally invert the lines passing through. Now, we use these two key bits to select one of four possible functions for each switch box output. 
One configuration produces the correct function, the others are randomly generated decoy functions of the switch box inputs. 

\begin{figure}[t]
  \centering
  \includegraphics[width=\columnwidth]{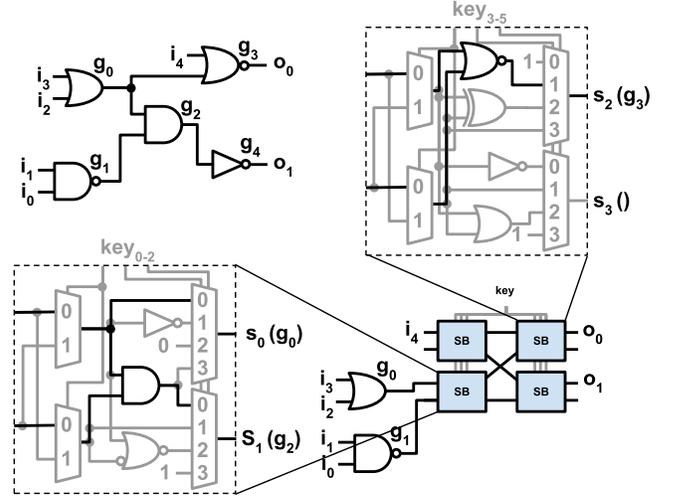}
  \caption{Diagram of circuit mapped to logic-enhanced Banyan network. The original circuit is shown top-left, the locked version bottom-right. The correct switch box function is highlighted in black, the decoy logic in gray.}
  \label{fig_switch}
\end{figure}

A diagram of the new technique is depicted in Fig. \ref{fig_switch}.
In this small example, a 4-input Banyan network is inserted. Using the switchbox outputs as reference points, gates from the original circuit are mapped to the Banyan network.
Switch box outputs $s_0$, $s_1$, and $s_2$ respectively map to gates $g_0$, $g_2$, and $g_3$. We show the internal logic of two of the switch boxes; the logic corresponding to the original circuit highlighted in black whereas the decoy logic is in gray. 
Input, $i_4$ feeds through the top-left switch box and gate $g4$ is mapped to the upper output of the bottom-right switch box. The network's un-mapped inputs and outputs are connected to the surrounding circuitry  

As the network size is increased, it incorporates a larger portion of the design. 
Since there is already a significant amount of reconfiguration, we forgo the use of LUTs. 
The intra-network logic prohibits the use of a simplified model for the network. The correct functionality is no longer just a permutation of the inputs to the network, but rather one of a very large space of functionalities dependent on nearly all the key bits. Additionally, the large amount of symmetry has been removed; while some corner case symmetry may remain, it will be highly complex to find and probably of little value to rule out. 

\begin{figure*}[t]
  \centering
  \includegraphics[width=\textwidth]{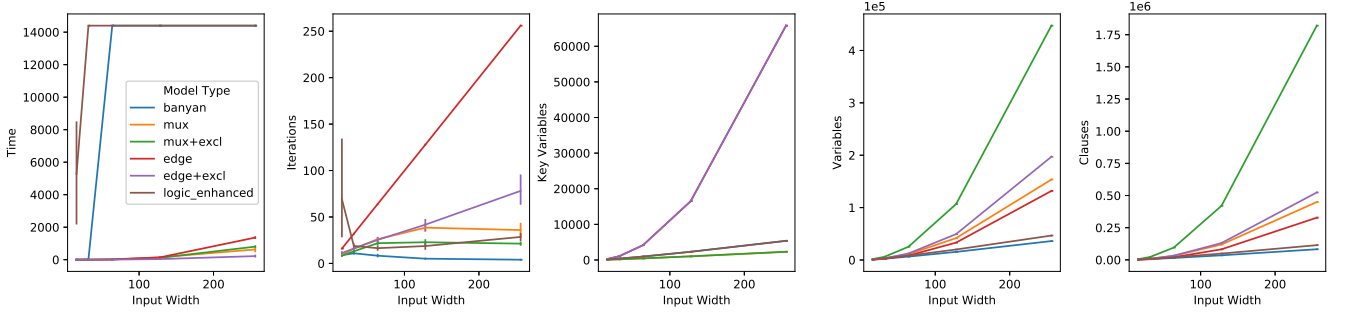}
  \caption{Comparison of encoding schemes for standalone Banyan network, n=10}
  \label{fig_enc}
\end{figure*}

\subsection{Insertion Algorithm}
While resistant to the modeling techniques, the insertion of the Banyan network is more complex than in Full-Lock. Now, gates from the original circuit must be mapped onto the structure of the Banyan network, instead of just being randomly selected. We automate this process to enable scalable exploration of the mapping solution space. To augment the ability to map onto the Banyan structure, we split all gates from the original circuit with three or more inputs into two input gates. 
We start with a Banyan network of the desired input width, $W$ and 
encode the problem of finding a mapping as a SAT instance through constraints that we specify below. 

The encoding uses a set of variables representing a mapping between an original gate $g$ and a banyan switch box output $s$. 
The switch box outputs provide a reference point within the Banyan network that naturally correspond to gate outputs in the original circuit.
For all pairs of gates in the original circuit and switch box outputs in the Banyan network, $(g,s) \in C\times B$, we create a mapping variable $m_{gs}$. The variable is true if gate $g$ is mapped to switch box output $s$.  
Over these variables, we encode constraints that ensure the amount of mapping is sufficient.
First we ensure at least W gates are mapped to the network. 
\begin{equation}
AtLeastW(\{\bigvee_{s \in B}m_{gs} :g \in C\})
\end{equation}
Then we encode that at most one gate is mapped per switch box output.
\begin{equation}
\bigwedge_{s\in B} AtMostOne(\{m_{gs} :g \in C\})
\end{equation}
We allow multiple switch box outputs to map to the same gate enabling the mapping of gates with fanout.
Finally, we prohibit any path directly feeding through from the network inputs to outputs, avoiding the simplest mappings. This is done by prohibiting a gate to be mapped to both the first and last layer of the Banyan network.
We show the encoding below where $B_i$ and $B_o$ are respectively the sets of switch box outputs in the first and last layers of the network. 
\begin{equation}
\bigwedge_{g\in C} \bigwedge_{s_i\in B_i} \bigwedge_{s_o\in B_o} AtMostOne(m_{gs_o},m_{gs_i})
\end{equation}

To maintain the structure of the circuit, we add constraints that enforce a correspondence between the connectivity of the mapped gates and the switch box outputs. If a gate is mapped to a switch box output, the fanin of the gate in the original circuit must be mapped to the fanin of the switch box (i.e. the switch box outputs from the preceeding network layer). Similarly, we also ensure that at least one of the gate's fanout is mapped to the fanout of the switch box. 
We allow an exception to this rule if the gate is simply fed through the switch box, which adds flexibility to the circuit structures which can be mapped. Note that here we are allowing feed through for a switch box, but prohibit it through the entire network.
More formally, $m_{gs}$ implies that every fanin of $g$ is mapped to the fanin of $s$ or, in the case of a feedthrough, $g$ itself is mapped to the fanin of $s$.
This encoding is shown below.
\begin{equation}
\bigwedge_{s\in B} \bigwedge_{g\in C} \bigwedge_{g_{f}\in fanin(g)} m_{gs} \to  \bigvee_{s_{f} \in fanin(s)} m_{g_{f}s_{f}} \vee m_{gs_{f}} 
\end{equation}
Additionally, $m_{gs}$ implies that at least one fanout of $g$ is mapped to the fanout of $s$ or, in the case of a feedthrough, $g$ is in the fanout of $s$. This encoding is shown below.
\begin{equation}
\bigwedge_{s\in B} \bigwedge_{g\in C} m_{gs} \to \bigvee_{s_{f} \in fanout(s)} \bigvee_{g_{f} \in fanout(g)\cup \{g\}}  m_{g_{f}s_{f}}
\end{equation}

This system of constraints is solved and the gates in the resulting mapping are inserted into their corresponding switch boxes and removed from the original circuit. The other MUX inputs are connected to randomly selected decoy functions of the switch box inputs. The network inputs and outputs are connected depending on which gates have been mapped to the first and last layer of switch boxes. It's important to emphasize that no intermediate connections are made to or from the network. 
Outputs with no mapping are randomly connected to remaining gates such that they have no impact on the logic of the system under the correct key.

\section{Attack Results}
In this section, we provide experimental data to show the effect of using the proposed modeling techniques. We step through each part of our modeling process, showing the incremental results from each. We then demonstrate the resistance of our proposed technique to the miter-based SAT attack. 

All attacks are run using a Python implementation of the miter-based SAT attack. The implementation uses PySAT's wrapper for the CDCL-based SAT solver CaDiCaL  \cite{imms-sat18,Cadical}. Additionally, the attack implementation uses incremental addition of constraints as proposed in \cite{CunxiYu2017ISRE}. Although Python is not as efficient as c or other low-level languages, most attack time is spent inside the SAT solver and thus the difference is negligible. The logic-enhanced Banyan locking implementation can be found in our repository\footnote{\url{https://github.com/jpsety/logic_enhanced_banyan_locking}}

Each attack has a timeout 4 of hours, an iteration count of 10, and is executed on a machine with 756GB RAM and 16 2.1GHz cores. The attacks are conducted in parallel while ensuring minimal contention for resources by allotting memory greater than the maximum usage of the largest instances to each run.

\subsection{Relaxed Model Comparison}
\label{sec_rel_enc}
We assess the impact of the model and encoding on attack run time for the standalone Banyan networks. 
We compare five model, encoding schemes as described in Section \ref{section_relaxed_models}, namely the original Banyan network model and encoding, and all combinations of MUX-based and edge-based encodings with the all-to-all and all-to-all exclusive models. We also compare our logic-enhanced Banyan locking. For each iteration we randomly select a new key. The data is shown in Fig. \ref{fig_enc}.
We report several dimensions: overall attack time, number of attack iterations, number of key variables, number of total variables, and number of clauses.

Immediately obvious is the grouping of attack times. The original and logic-enhanced Banyan networks respectively timeout at input widths of 64 and 16.  
Whereas the attack time of all proposed model-encoding pairs is significantly less, highlighting the impact of the improved models. 
The number of iterations completed for the original and logic-enhanced Banyan models remain low, a testament to the hardness of the problems. 
While the edge-based encoding has significantly more key variables, the overall variable and clause counts remain close to the MUX-based encoding's values. Both encoding schemes result in larger formulations than the original Banyan network, however are significantly easier to solve. 
The MUX-based all-to-all exclusive encoding stands out as by far the largest relaxed encoding.
 
\begin{figure}[t]
  \centering
  \includegraphics[width=\columnwidth]{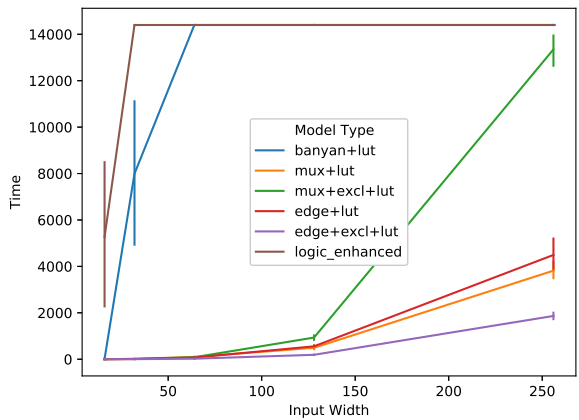}
  \caption{Attack time comparison of encoding schemes for Banyan network with 2-input LUTs connected to the outputs, n=10}
  \label{fig_enc_lut}
\end{figure}

We then add the output LUTs to the Full-Lock network models and repeat the attacks, reporting just the execution time in Fig. \ref{fig_enc_lut}. Again, for comparison we show our proposed logic-enhanced Banyan locking scheme. 
Here we can clearly discern that the edge-based all-to-all encoding performs the best, quickly terminating even with an input width of 256. For reference, this scenario would require the original Banyan network 5,889 keys to implement (two thirds of which are dedicated to inversions as in Fig. \ref{fig_fulllock_diag}). 

\subsection{Effect of Symmetry Breaking}
\begin{figure}[t]
  \centering
  \includegraphics[width=\columnwidth]{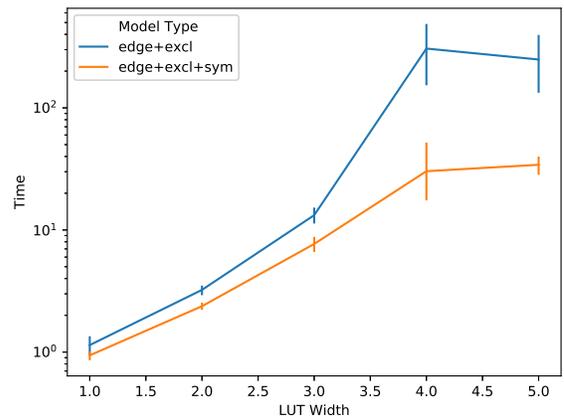}
  \caption{Comparison of attack time at network input width of 32 between encodings with and without LUT symmetry breaking, n=10}
  \label{fig_sym}
\end{figure}

Next, we demonstrate the impact of the LUT symmetry breaking on attack time. Since the amount of symmetry scales with LUT input size, we sweep this parameter and hold the network width fixed at 32. The resulting attack times with and without symmetry are shown in Fig. \ref{fig_sym}. As the LUT width increases, the advantage of symmetry breaking grows exponentially. At a LUT input width of 5, the difference in attack time about an order of magnitude. This result is consistent with the difference in potential solutions reported in section \ref{sec_sym}. 
\begin{figure*}[t]
  \centering
  \includegraphics[width=\textwidth]{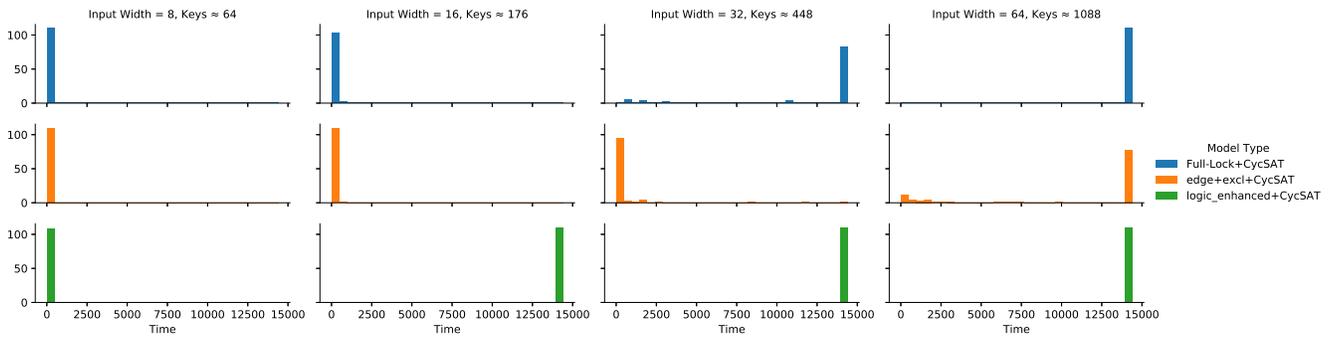}
  \caption{SAT-based attack time for ISCAS85 circuits locked with Full-Lock and logic-enhanced Banyan locking schemes}
  \label{fig_fulllock}
\end{figure*}

\subsection{Full-Lock and Logic-Enhanced Banyan Attack Results}
Finally, the proposed techniques are together applied to Full-Lock.
We use the best encoding from the previous results: an edge-based, all-to-all exclusive with symmetry breaking along with the CycSAT acyclic constraints \cite{Zhou2017CycSAT:Encryptions}. This is compared to the attack model outlined by the Full-Lock authors: the full Banyan network model alongside the CycSAT acyclic constraints. 
We also demonstrate our proposed solution, logic-enhanced Banyan locking.
All techniques are run on the ISCAS85 benchmark circuits, sweeping the network input width from 8 to 64. The corresponding key widths range is around 64 to 1088. The results are shown in Fig. \ref{fig_fulllock}.

The Full-Lock run times show a trend that mostly agrees with the original results with the exception that some circuits at 32-input width are deobfuscated. These improved results are likely due to the use of a different SAT solver. Our relaxed model shows run times several orders of lower than the original paper. Most circuits at 32-input width (around 448 key bits) are deobfuscated in seconds, with some outliers taking minutes, clearly demonstrating the effectiveness of these techniques. As the input width scales to 64, many circuits are still deobfuscated, but the majority take longer than our 4-hour timeout. 

The logic-enhanced Banyan scheme provides a significantly better ratio of key bits to attack time than the Full-lock predecessor. At an input width of 16, the attack times out for all circuits. While this is good, we do not suggest that such small input widths are viable locking techniques as simple enumeration attack schemes may easily deobfuscate them. 

\section{Discussion}
Our experiments show that these modeling techniques are highly effective. The actual attack times for the original Full-Lock implementation are unknown, we just have a lower bound. With this, we can claim that the techniques have decreased attack times by \textit{at least} several orders of magnitude. 
Important to the success of our strategy was trying different models of the system. 

Several concerns remain unexplored.
First, the amount of corruption produced by the remaining keys after the miter-based SAT attack has been run for some time. It is common to only report attack run times, however, the keys that can be obtained at timeout may be very close to a correct solution. Understanding the trend in remaining corruption is critical information for a given locking scheme. 
Also, an attacker can often specify constraints on the keys that the correct solution must respect. 
One applicable example is timing constraints, wherein every valid key must produce a circuit with a critical path less than the period. 
Just like the acyclic constraints that are necessary for the attack to complete, timing constraints could rule out significant portions of the key space. Importantly, like our modeling techniques, encoding has a large impact on effectiveness of a constraint. 

More specific to our proposed technique, an overhead analysis has not been conducted. It is likely that more must be done to make this a viable solution for high speed designs. 
Towards that end, a parameter exploration of similar structures may produce increased attack resistance. Our initial implementation maintained the amount of keys from Full-Lock. Varying the amount of keys, connectivity of the network, or decoy logic selection may produce significantly overhead and attack resistance results.

\section{Conclusion}
We have proposed two widely-applicable modeling techniques that can substantially decrease the attack time for logic locking schemes. Any locking scheme that has human-comprehensible regularities is potentially vulnerable to these techniques. 
We have demonstrated the application of these techniques on a state-of-the-art locking scheme, Full-Lock. The experiments show many orders of magnitude decrease in attack time compared to previously reported results. 
In general, these modeling techniques are essential considerations in any logic locking technique. 

Additionally, we have described logic-enhanced Banyan locking, an extension to the Full-Lock method, that appears to be resistant to these modeling techniques.
We demonstrated promising initial attack results, showing that structure of the Banyan network combined with randomly selected decoy logic is not only a mechanism of resisting these techniques, but also harder for the CDCL-based SAT solver used. 
Of course, this resistance may change with some additional insight or varied attack strategies.

\section*{Acknowledgments}
This work was supported in part by the Defense Advanced Research Projects Agency under contract FA8750-17-1-0059 “Obfuscated
Manufacturing for GPS (OMG)”, Honeywell Federal Manufacturing \& Technologies, LLC under contract A023646, and NSF grant CCF-1618574.

\bibliographystyle{ACM-Reference-Format}
\bibliography{sat}

\end{document}